\documentclass{IEEEtran}

\usepackage[noadjust]{cite}

\usepackage[T1]{fontenc}
\usepackage{newtxtext}
\usepackage[cmintegrals]{newtxmath}
\usepackage{bm}
\usepackage{amsmath,amsfonts}

\usepackage{graphicx}
\usepackage[caption=false,font=footnotesize]{subfig}
\usepackage{array}
\usepackage{float}
\usepackage{wrapfig} 

\usepackage{xcolor}
\usepackage{textcomp}
\usepackage{url}
\usepackage{balance}
\usepackage{algorithm}
\usepackage{algpseudocode}

\usepackage[hypertexnames=true, bookmarks=true, bookmarksnumbered=true,
pdfpagemode=UseOutlines, plainpages=false, pdfpagelabels=true,
colorlinks=true, linkcolor=black, citecolor=black, urlcolor=black]{hyperref}

\begin{document}

\title{Modeling the Unlicensed Band Allocation for LAA With Buffering Mechanism}

\author{Po-Heng Chou,~\IEEEmembership{Member,~IEEE} \vspace{-0.4in}

\thanks{Manuscript received December, 14, 2018; accepted January, 5, 2019.}
\thanks{P.-H. Chou is with the Graduate Institute of Communication Engineering (GICE), National Taiwan University (NTU), Taipei 10617, Taiwan (e-mail:d00942015@ntu.edu.tw).}
\thanks{This work was supported in part by the Ministry of Science and
Technology (MOST) of Taiwan under Grant MOST 103-2218-E-002-032, 104-3115-E-002-004, 105-2218-E-002-013, and 106-2218-E-002-029.}
}

\markboth{IEEE COMMUNICATIONS LETTERS,~VOL.~23, NO.~3, Mar.~2019}
{}

\maketitle

\begin{abstract}
In this letter, we propose an analytical model and conduct simulation experiments to study listen-before-talk-based unlicensed band allocation with the buffering mechanism for the License-Assisted Access (LAA) packets in the heterogeneous networks. In such a network, unlicensed band allocation for LAA and Wi-Fi is an important issue, which may affect the quality of service for both systems significantly. We evaluate the performance of these unlicensed band allocations in terms of the acceptance rate of both LAA and Wi-Fi packets. This letter provides the guidelines for designing the channel occupation phase and buffer threshold of the LAA systems.
\end{abstract}


\IEEEpeerreviewmaketitle

\section{Introduction}
\label{sect:introduction}
Diversity of wireless Internet applications has driven high demand for more wireless bandwidth in the next generation of mobile networks. Utilization of the unlicensed band has been considered as one of the promising technologies to have more wireless bandwidth. License-Assisted Access (LAA)~\cite{3GPP2015} is proposed by the 3rd
Generation Partnership Project (3GPP) to have LTE
utilize the unlicensed band that is also used by the incumbent
systems (e.g., Wi-Fi).
The LAA adopts the Listen Before Talk (LBT) mechanism, where the LTE base station (eNB)
periodically senses the target unlicensed band to determine whether the unlicensed band is idle or not, and begins the data
transmission upon if the target unlicensed band is idle.
Most of the previous studies that focused on the fairness between the LAA and Wi-Fi.
However, Wi-Fi and LAA compete for shared resources, and due to the inherent properties of the CSMA protocol family, strict delay constraints cannot be met. And the LBT mechanism is fundamentally different from the CSMA/CA protocol,~\cite{Maule2018}, suggesting that the parameters that identify the fairness of resource allocation may not be trivial. Since the LAA eNB can be granted full control over the shared spectrum, competition between LTE and Wi-Fi may not be required to achieve fairness. Therefore, the LAA eNB can manage resource allocation between Wi-Fi and LTE instead of competing with Wi-Fi to divide them in a fair manner while tracking the number of concurrent LTE packets and the status of unlicensed bands for the LAA eNB.
The previous works~\cite{Tang2018,Song2016,Chen2015} considered the interactions between the LAA and the Wi-Fi networks by utilizing two Markov chains separately.
However, the interaction between the Markov model for LAA and that for Wi-Fi was modeled by only the collision probability,
which may not properly reflect the interaction in the real systems.
To address this issue, in this letter, we model the interaction between LAA and Wi-Fi by using a global system state.
Furthermore, the MAC layer performance studies in the previous works did not consider the traffic model in the application layer,
so the studies may not reflect the situation in the real system.

In this letter, we consider both the application traffic model (based on the
3GPP FTP traffic model~\cite{3GPP36.814}) in the analytical model and the simulation model.
This letter proposes an analytical model to study the coexisting issue for LAA small cell operating on the unlicensed band with Wi-Fi heterogeneous networks.
This model quantifies the length of the channel occupancy phase of the LAA small cell, which is referred to as the ON / OFF service rate. Obviously, the shorter the length of the channel occupancy phase, the smaller the impact on the Wi-Fi system. However, reducing the length of the channel occupancy phase may also reduce LAA system performance in terms of dropping probability. Therefore, we also quantify the LAA packet dropping probability to check for penalties due to unlicensed band allocation to the LAA/Wi-Fi heterogeneous network. Due to the complex ON/OFF behavior of the LAA small battery, the proposed analytical model may not be able to capture the ON/OFF behavior of the LAA small battery under certain conditions. In order to release these constraints of the analytical model, we also conducted simulation experiments. Although~\cite{Song2016, Chen2015} also proposed mathematical frameworks, they do not consider the simulation model to verify the numerical results. We note that the performance of unlicensed band allocation is primarily dependent on the length of the sensing phase/channel occupancy phase and the Wi-Fi arrival rate. In addition, our research shows that with proper parameter settings, the buffering mechanism can reduce the impact of Wi-Fi without significantly increasing the probability of LAA packet drop. The results of this work can be used as a guiding principle for implementing LAA small batteries.

\section{LBT-based Allocation with Buffering}
\label{sect:allocation}

\begin{figure}[t]
\centering
\includegraphics[width=.3\textwidth]{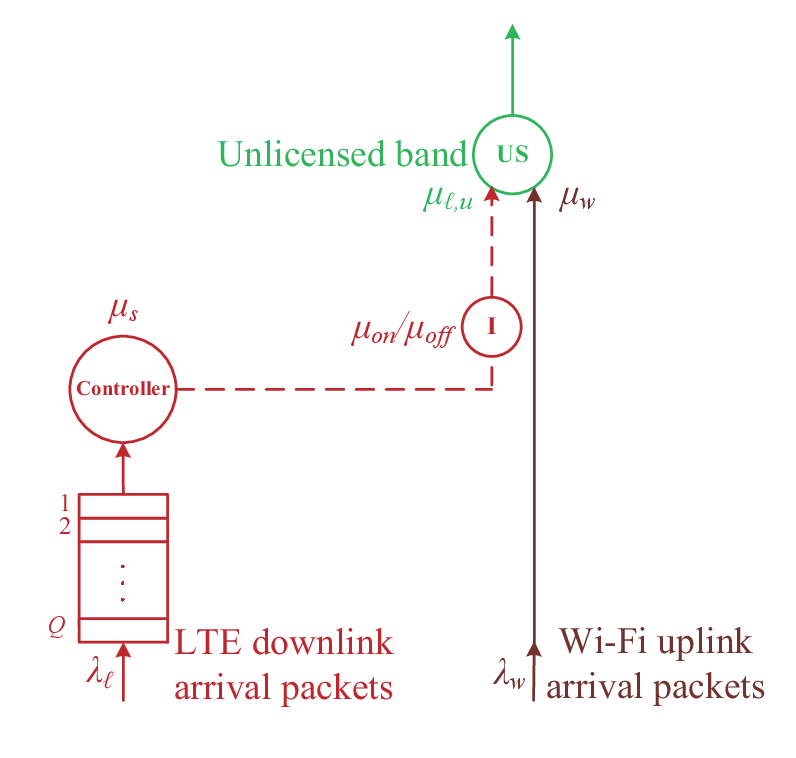}
\caption{The queueing model of LAA coexisting with Wi-Fi heterogeneous networks.}
\label{LAA/Wifi_Queuing}
\end{figure}

Fig.~\ref{LAA/Wifi_Queuing} shows the queuing model for the LAA and Wi-Fi coexisting system.
If the FIFO queue is not empty,  the controller will execute the sensing phase and the occupancy phase alternatively. The length of the sensing phase and occupancy phase is determined by the {\bf timer} of the sensing phase and occupancy phase, respectively. If the unlicensed server is free, the controller sets the occupancy phase to the ON state. Otherwise, the controller sets the occupancy phase to the OFF state.
If the LAA small cell requires more data transmission time, it will re-execute the CCA for the next channel occupancy phase and reallocate the resources for downlink transmission.
When an LAA packet arrives, the LAA small cell checks the status of the FIFO queue and the unlicensed channel pool. We consider three cases:\\
Case 1 :$Q_{f} \geq Q_{\theta}$ (where $Q_{\theta}$ is a threshold for buffering control). The LAA packet will stay in the FIFO queue.\\
Case 2 : $Q_{f} < Q_{\theta}$ and $D_{f} > 0$ The LAA small cell assigns one unlicensed channel to serve the LAA packet.\\
Case 3 : $Q_{f} = 0$ and $D_{f} = 0$ The LAA packet is dropped.

\section{Analytical Model}
\label{sect:analytical}

In this section, we propose an analytical model to investigate the performance of the LBT-based allocation with the buffering mechanism as shown in Fig.~\ref{LAA/Wifi_Queuing}.
The periods of the sensing phase of the LAA small cell are assumed to have an exponential distribution with mean $1/\mu_s$.
The LAA packet arrivals and Wi-Fi packet arrivals to a cell are assumed to form Poisson processes with rates $\lambda_{\ell}$ and $\lambda_{w}$, respectively.
The occupancy phase of the LAA packet $t_{\ell, u}$ on an unlicensed channel is assumed to have exponential distribution with the density function $f_{\ell, u}(t_{\ell, u})=\mu_{\ell, u}e^{-\mu_{\ell, u}t_{\ell, u}}$ and
$E[t_{\ell, u}] =1/\mu_{\ell, u}$. The Wi-Fi packet transmission time $t_{w}$ on an unlicensed channel is assumed to have an exponential distribution with the density function
$f_{w}(t_{w})=\mu_{w}e^{-\mu_{w}t_{w}}$ and
$E[t_{w}] =1/\mu_{w}$.
Note that in the real world, the LAA or Wi-Fi packet inter-arrival times and packet transmission times may not be exponentially distributed.
Having exponential assumptions, our analytic models serve two purposes.
First, the exponential distribution can provide the mean value analysis.
Second, the analytic model is for the validation of the simulation experiments, based on which we can study the effects of different distributions.

\begin{figure}[t]
\centering
\includegraphics[width=.5\textwidth]{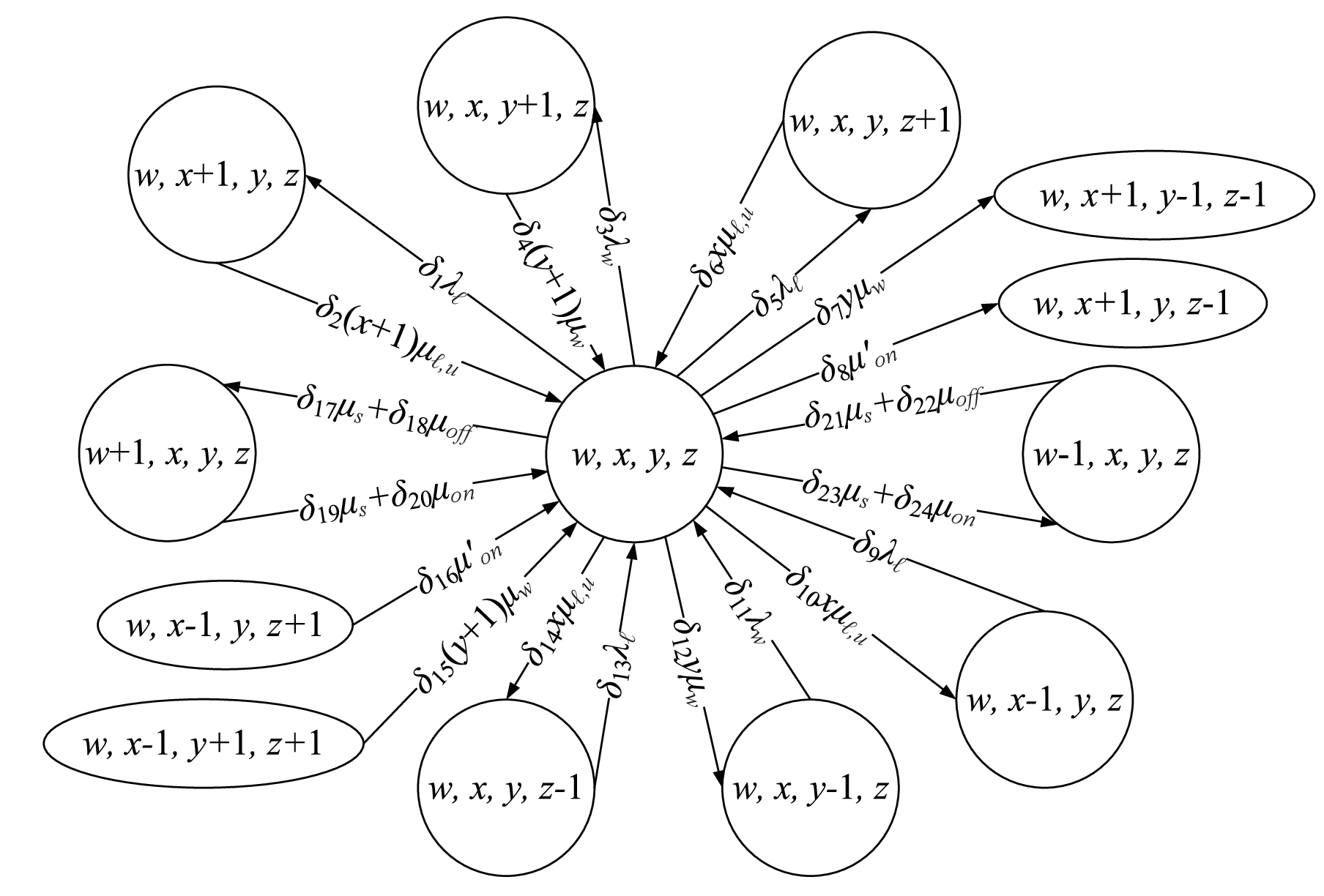}
\caption{The state transition diagram for the LBT-based allocation with the buffer mechanism scheme.}
\label{Basic_Queuing}
\end{figure}

The behavior of the LAA and Wi-Fi coexisting network can be modeled by a Markov process with the system state $\mathbf{n}= (w, x, y, z)$.
$w$ denotes the status of an unlicensed channel, where $w=1$ implies that the LAA system is performing the CCA, and $w=2$ implies that the unlicensed channel is at the {OFF} state and the {ON} state, respectively.
To solve the system, we employ the finite Markov chain approximation by limiting the number of unlicensed band servers to one.
Note that the proposed analytical model can be easily extended for multiple unlicensed band servers.
This can be achieved by adding a state for each unlicensed band server.
Due to page limitations, we do not include the scenario of multiple servers in this letter.
$x = 1$ and $x = 0$ imply that there is an LAA packet being served by the unlicensed channel and no LAA packet being served by the unlicensed channel, respectively.
$y = 1$ and $y = 0$ imply that there is a Wi-Fi packet being served by the unlicensed channel and no Wi-Fi packet being served by the unlicensed channel, respectively.
$z$ denotes the number of LAA  packets buffered in the FIFO queue.
We use the Zachary-Kelly model~\cite{Zachary1991, Kelly1991} together
with an iterative algorithm to derive the LAA and Wi-Fi packet
dropping probability (i.e., $P_{b, \ell}$ and $P_{b, w}$). The state space of the Markov process for the LBT-based allocation with the buffering mechanism
\begin{align}
\mathbf{S} = \big\{&\mathbf{n} |  w \in \{0, 1, 2\}, x \in \{0, 1\}, y \in \{0, 1\},\notag\\
&z \in \{0, 1, \ldots , Q\}, Q_{\theta} \in \{1, \ldots , Q-1\} \big\}.
\end{align}

In addition, we let $\pi_{w,x,y,z}$ be the steady state probability
for state $(w, x, y, z)$, where $\pi_{w,x,y,z} = 0$ if state
$(w,x,y,z) \notin \mathbf{S}$. For all legal states $(w,x,y,z)
\in \mathbf{S}$, $\sum_{w,x,y,z \in
\mathbf{S}}\pi_{w,x,y,z} = 1$.

The state transition diagram for an LAA coexisting with Wi-Fi
heterogeneous network is illustrated as~Fig.~\ref{Basic_Queuing} in
the previous section. The transitions of the Markov process are
described as follows. If state $(w, x, y, z) \in \mathbf{S}$,
the following transitions should be considered.

The balance equations for this process is $\pi_{w, x, y, z} = \frac{\varepsilon_{A}}{\varepsilon_{B}}$, where
\begin{align}
\varepsilon_{A} = &\; \delta_{2}(x+1)\mu_{\ell, u}\pi_{w, x+1, y, z} + \delta_{4}(y+1)\mu_{w}\pi_{w, x, y+1, z} \notag\\
&+ \delta_{6}x\mu_{\ell, u}\pi_{w, x, y, z+1}+ \delta_{9}\lambda_{\ell}\pi_{w, x-1, y, z} + \delta_{11}\lambda_{w}\pi_{w, x, y-1, z}\notag\\
&+ \delta_{13}\lambda_{\ell}\pi_{w, x, y, z-1}+ \delta_{15}(y+1)\mu_{w}\pi_{w, x-1, y+1, z+1} \notag\\
&+ \delta_{16}\mu_{on}'\pi_{w, x-1, y, z+1}+ (\delta_{19}\mu_{s} + \delta_{20}\mu_{on})\pi_{w+1, x, y, z}\notag\\
&+ (\delta_{21}\mu_{s} + \delta_{22}\mu_{off})\pi_{w-1, x, y, z},
\end{align}
and
\begin{align}
\varepsilon_{B} =& \; \delta_{1}\lambda_{\ell} + \delta_{3}\lambda_{w} + \delta_{5}\lambda_{\ell} + \delta_{7}y\mu_{w} + \delta_{8}\mu_{on}' + \delta_{10}x\mu_{\ell, u}\notag\\
&+ \delta_{12}y\mu_{w} + \delta_{14}x\mu_{\ell, u} + \delta_{17}\mu_{s} + \delta_{18}\mu_{off}+ \delta_{23}\mu_{s}\notag\\ &+ \delta_{24}\mu_{on}.
\end{align}

$\delta_{1}, \delta_{2}, \ldots , \delta_{24}$ are obtained in the following equations:

\begin{itemize}
\item[\textbf{1)}] If an LAA packet arrives when the process is at state $(w, x, y, z) \in \mathbf{S}$, the LAA state is ON and the queue is empty, one unlicensed channel is assigned to this LAA packet. Define
\begin{align}
\delta_{1}= \left\{
\begin{array}{ll}
1, & w=2,~0 \leq x < D-y,~z=0,\\
& \textrm{and} \; (w, x, y, z) \in \mathbf{S}\\
0, & \textrm{otherwise}.
\end{array}\right.
\label{UTA_1}
\end{align}
The process moves from $(w, x, y, z)$ to $(w, x+1, y, z)$
with rate $\delta_{1}\lambda_{\ell}$.

\item[\textbf{2)}] If the transmission for an LAA packet with one allocated unlicensed channel completes at state $(w, x+1, y, z) \in \mathbf{S}$, then the channel is released. Define
\begin{align}
\delta_{2}= \left\{
\begin{array}{ll}
1, & (w=2,~0 < x \leq D-y,~z=0)\\
&\cup\; (w \neq 2,~0 < x \leq D-y),\\
& \textrm{and} \; (w, x+1, y, z) \in \mathbf{S}\\
0, & \textrm{otherwise}.
\end{array}\right.
\label{UTA_2}
\end{align}
Transition: $(w, x+1, y, z) \to (w, x, y, z)$ with rate $\delta_{2}(x+1)\mu_{\ell, u}$.

\item[\textbf{3)}] If a Wi-Fi packet arrives at $(w, x, y, z) \in \mathbf{S}$, one unlicensed channel is assigned. Define
\begin{align}
\delta_{3}= \left\{
\begin{array}{ll}
1, & 0 \leq y < D-x,\textrm{and} \; (w, x, y, z) \in \mathbf{S}\\
0, & \textrm{otherwise}.
\end{array}\right.
\label{UTA_3}
\end{align}
Transition: $(w, x, y, z) \to (w, x, y+1, z)$ with rate $\delta_{3}\lambda_{w}$.

\item[\textbf{4)}] If the transmission for a Wi-Fi packet with one allocated unlicensed channel completes at $(w, x, y+1, z) \in \mathbf{S}$, the channel is released. Define
\begin{align}
\delta_{4}= \left\{
\begin{array}{ll}
1, & 0 < y+1 \leq D-x,\\
& \textrm{and} \; (w, x, y+1, z) \in \mathbf{S}\\
0, & \textrm{otherwise}.
\end{array}\right.
\end{align}
Transition: $(w, x, y+1, z) \to (w, x, y, z)$ with rate $\delta_{4}(y+1)\mu_{w}$.

\item[\textbf{5)}] If an LAA packet arrives at $(w, x, y, z) \in \mathbf{S}$ where no unlicensed channel is free or the LAA state is not ON, and the queue is not full, the packet is buffered. Define
\begin{align}
\delta_{5}= \left\{
\begin{array}{ll}
1, & (w=2,~0 \leq z < Q,~x+y=D)\\
&\cup\; (w \neq 2,~0 \leq z < Q),\\
& \textrm{and} \; (w, x, y, z) \in \mathbf{S}\\
0, & \textrm{otherwise}.
\end{array}\right.
\end{align}
Transition: $(w, x, y, z) \to (w, x, y, z+1)$ with rate $\delta_{5}\lambda_{\ell}$.

\item[\textbf{6)}] If an LAA transmission completes at $(w, x, y, z+1) \in \mathbf{S}$ and the LAA state is ON, one queued LAA packet is served. Define
\begin{align}
\delta_{6}= \left\{
\begin{array}{ll}
1, & w=2,~0 < z+1 \leq Q,~x+y = D,\\
& \textrm{and} \; (w, x, y, z+1) \in \mathbf{S}\\
0, & \textrm{otherwise}.
\end{array}\right.
\end{align}
Transition: $(w, x, y, z+1) \to (w, x, y, z)$ with rate $\delta_{6}x\mu_{\ell, u}$.

\item[\textbf{7)}] If a Wi-Fi transmission completes at $(w, x, y, z) \in \mathbf{S}$ and the LAA state is ON, one queued LAA packet is served. Define
\begin{align}
\delta_{7}= \left\{
\begin{array}{ll}
1, & w=2,~y > 0,~z > 0,~x+y = D,\\
& \textrm{and} \; (w, x, y, z) \in \mathbf{S}\\
0, & \textrm{otherwise}.
\end{array}\right.
\end{align}
Transition: $(w, x, y, z) \to (w, x+1, y-1, z-1)$ with rate $\delta_{7}y\mu_{w}$.

\item[\textbf{8)}] When an unlicensed channel is free, if the LAA state switches to ON and the queue is nonempty, one LAA packet is served. Define
\begin{align}
\delta_{8}= \left\{
\begin{array}{ll}
1, & w = 2,~y = 0,~z > 0,\\
&0 \leq x < D-y,\\
& \textrm{and} \; (w, x, y, z) \in \mathbf{S}\\
0, & \textrm{otherwise}.
\end{array}\right.
\label{UTA_8}
\end{align}
Transition: $(w, x, y, z) \to (w, x+1, y, z-1)$ with rate $\delta_{8}\mu_{on}'$, where $\mu_{on}' = 10\mu_{on}$.
\end{itemize}

    The following $\delta_{9}$, $\delta_{10}$,\ldots , $\delta_{16}$ terms are the complementary (reverse or symmetric) cases of $\delta_{1}$, $\delta_{2}$,\ldots , $\delta_{8}$. They appear in the balance equations alongside \(\pi_{w,\,x\pm 1,\,y\pm 1,\,z\pm 1}\).

\begin{itemize}
\item[\textbf{9)}] (Complement of \(\delta_1\)) LAA arrival that \emph{would} start service when \(w=2\), \(z=0\):
\begin{align}
\delta_{9}=\left\{\!
\begin{array}{ll}
1, & w=2,\; 0\le x-1 < D-y,\\
& z=0, \textrm{and} \; (w,x-1,y,z)\in\mathbf{S}\\[1pt]
0, & \text{otherwise}.
\end{array}\right.
\end{align}
Transition used in the balance term: \((w,x-1,y,z)\!\to\!(w,x,y,z)\) at rate \(\delta_9\,\lambda_{\ell}\).

\item[\textbf{10)}] (Complement of \(\delta_2\)) LAA service completion:
\begin{align}
\delta_{10}=\left\{\!
\begin{array}{ll}
1, & (w=2,\; 0<x\le D-y,\; z=0)\;\\
&\cup\;(w\neq 2,\;0<x\le D-y),\\
& \textrm{and} \; (w,x,y,z)\in\mathbf{S}\\[1pt]
0, & \text{otherwise}.
\end{array}\right.
\end{align}
Transition: \((w,x,y,z)\!\to\!(w,x-1,y,z)\) at rate \(\delta_{10}\,x\,\mu_{\ell,u}\).

\item[\textbf{11)}] (Complement of \(\delta_3\)) Wi-Fi arrival that \emph{would} start service:
\begin{align}
\delta_{11}=\left\{\!
\begin{array}{ll}
1, & 0\le y-1 < D-x,\\
& \textrm{and} \; (w,x,y-1,z)\in\mathbf{S}\\[1pt]
0, & \text{otherwise}.
\end{array}\right.
\end{align}
Transition: \((w,x,y-1,z)\!\to\!(w,x,y,z)\) at rate \(\delta_{11}\,\lambda_w\).

\item[\textbf{12)}] (Complement of \(\delta_4\)) Wi-Fi service completion:
\begin{align}
\delta_{12}=\left\{\!
\begin{array}{ll}
1, & 0<y\le D-x,\\
& \textrm{and} \; (w,x,y,z)\in\mathbf{S}\\[1pt]
0, & \text{otherwise}.
\end{array}\right.
\end{align}
Transition: \((w,x,y,z)\!\to\!(w,x,y-1,z)\) at rate \(\delta_{12}\,y\,\mu_w\).

\item[\textbf{13)}] (Complement of \(\delta_{5}\)) LAA arrival that joins the queue:
\begin{align}
\delta_{13}=\left\{\!
\begin{array}{ll}
1, & (w=2,\; 0\le z-1 < Q,\; x+y=D)\\
&\cup\;(w\neq 2,\; 0\le z-1<Q),\\
&\textrm{and} \;  (w,x,y,z-1)\in\mathbf{S}\\[1pt]
0, & \text{otherwise}.
\end{array}\right.
\end{align}
Transition: \((w,x,y,z-1)\!\to\!(w,x,y,z)\) at rate \(\delta_{13}\,\lambda_{\ell}\).

\item[\textbf{14)}] (Complement of \(\delta_{6}\)) Dequeue one LAA upon LAA completion when \(w=2\):
\begin{align}
\delta_{14}=\left\{\!
\begin{array}{ll}
1, & w=2,\; 0<z\le Q,\; x+y=D,\\
&\textrm{and} \; (w,x,y,z)\in\mathbf{S}\\[1pt]
0, & \text{otherwise}.
\end{array}\right.
\end{align}
Transition: \((w,x,y,z)\!\to\!(w,x,y,z-1)\) at rate \(\delta_{14}\,x\,\mu_{\ell,u}\).

\item[\textbf{15)}] (Complement of \(\delta_{7}\)) Dequeue one LAA upon Wi-Fi completion when \(w=2\):
\begin{align}
\delta_{15}=\left\{\!
\begin{array}{ll}
1, & w=2,\; y+1>0,\; z+1>0,\\
&x+y=D,\\
&\textrm{and} \;  (w,x-1,y+1,z+1)\in\mathbf{S}\\[1pt]
0, & \text{otherwise}.
\end{array}\right.
\end{align}
Transition: \((w,x-1,y+1,z+1)\!\to\!(w,x,y,z)\) at rate \(\delta_{15}\,(y+1)\,\mu_w\).

\item[\textbf{16)}] (Complement of \(\delta_{8}\)) Switch to ON and immediately serve from queue (reverse form):
\begin{align}
\delta_{16}=\left\{\!
\begin{array}{ll}
1, & w=2,\; y=0,\; z+1>0,\\
& 0\le x-1< D-y,\\
&\textrm{and} \;  (w,x-1,y,z+1)\in\mathbf{S}\\[1pt]
0, & \text{otherwise}.
\end{array}\right.
\end{align}
Transition: \((w,x-1,y,z+1)\!\to\!(w,x,y,z)\) at rate \(\delta_{16}\,\mu_{on}'\) (where \(\mu_{on}'=10\,\mu_{on}\)).
\end{itemize}

The state transitions of the LAA small cell ($w \in \{0 (\textrm{OFF}), 1 (\textrm{Sensing}), 2 (\textrm{ON})\}$) are described as follows, and $\delta_{17}$, $\delta_{18}$, \ldots
, $\delta_{24}$ are obtained in the following equations:

\begin{itemize}
\item[\textbf{9)}] When the LAA state is Sensing, if the unlicensed channel is free and the queue length exceeds $Q_{\theta}$, the state switches to ON. Define
\begin{align}
\delta_{17} = \left\{
\begin{array}{ll}
1, & w=1,~x=0,~y=0,~z>Q_{\theta},\\
& \textrm{and} \; (w, x, y, z) \in \mathbf{S}\\
0, & \textrm{otherwise}.
\end{array}\right.
\label{UTAB_17}
\end{align}
Transition: $(w, x, y, z) \to (w+1, x, y, z)$ with rate $\delta_{17}\mu_{s}$.

\item[\textbf{10)}] When the LAA state is OFF, if $z>Q_{\theta}$, the state switches to Sensing. Define
\begin{align}
\delta_{18} = \left\{
\begin{array}{ll}
1, & w=0,~z>Q_{\theta}, \\
&\textrm{and} \;(w, x, y, z) \in \mathbf{S}\\
0, & \textrm{otherwise}.
\end{array}\right.
\label{UTAB_18}
\end{align}
Transition: $(w, x, y, z) \to (w+1, x, y, z)$ with rate $\delta_{18}\mu_{off}$.

\item[\textbf{11)}] When the LAA state is Sensing, if the channel is occupied by a Wi-Fi packet \emph{or} $z<Q_{\theta}$, the state switches to OFF. Define
\begin{align}
\delta_{19} = \left\{
\begin{array}{ll}
1, & (w+1=1,~x=0,~y=1)\\
& \cup~(w+1=1,~z< Q_{\theta}),\\
& (w+1, x, y, z) \in \mathbf{S}\\
0, & \textrm{otherwise}.
\end{array}\right.
\label{UTAB_19}
\end{align}
Transition: $(w+1, x, y, z) \to (w, x, y, z)$ with rate $\delta_{19}\mu_{s}$.

\item[\textbf{12)}] When the LAA state is ON, it switches to Sensing. Define
\begin{align}
\delta_{20} = \left\{
\begin{array}{ll}
1, & w+1=2, \\
& \textrm{and} \; (w+1, x, y, z) \in \mathbf{S}\\
0, & \textrm{otherwise}.
\end{array}\right.
\end{align}
Transition: $(w+1, x, y, z) \to (w, x, y, z)$ with rate $\delta_{20}\mu_{on}$.
\end{itemize}


The following $\delta_{21}$, $\delta_{22}$,\ldots , $\delta_{24}$ are the symmetric counterparts of $\delta_{17}$, $\delta_{18}$,\ldots , $\delta_{20}$ that appear in the balance terms with \(\pi_{w-1,\cdot}\).

\begin{align}
\delta_{21} &= \left\{\!
\begin{array}{ll}
1, & w-1=1,\; x=0,\; y=0,\; z>Q_{\theta},\\
&\textrm{and} \; (w-1,x,y,z)\in\mathbf{S}\\[1pt]
0, & \text{otherwise},
\end{array}\right.\\[3pt]
\delta_{22} &= \left\{\!
\begin{array}{ll}
1, & w-1=0,\; z>Q_{\theta},\\
&\textrm{and} \; (w-1,x,y,z)\in\mathbf{S}\\[1pt]
0, & \text{otherwise},
\end{array}\right.\\[3pt]
\delta_{23} &= \left\{\!
\begin{array}{ll}
1, & (w=1,\; x=0,\; y=1)\\
&\cup\;(w=1,\; z<Q_{\theta}),\\
&\textrm{and} \;(w,x,y,z)\in\mathbf{S}\\[1pt]
0, & \text{otherwise},
\end{array}\right.\\[3pt]
\delta_{24} &= \left\{\!
\begin{array}{ll}
1, & w=2,\; (w,x,y,z)\in\mathbf{S}\\[1pt]
0, & \text{otherwise}.
\end{array}\right.
\end{align}

\noindent
\emph{Remarks.} 
(i) $\delta_{9}$, $\delta_{10}$,\ldots , $\delta_{16}$ mirror the allocation/completion/queueing events of $\delta_{1}$, $\delta_{2}$,\ldots , $\delta_{8}$ so that the global balance equations correctly couple \(\pi_{w,x,y,z}\) with neighboring states.  
(ii) $\delta_{21}$, $\delta_{22}$,\ldots , $\delta_{24}$ provide the symmetric (index \(w\pm1\)) gating for the OFF/Sensing/ON transitions used by the LBT controller and the buffer threshold \(Q_\theta\).

When an LAA packet arrives at the states where the queue is
full, and there is no free unlicensed server (i.e.,
$D_{f}=0, Q_{f}=0$), this LAA packet will be dropped. On
the other hand, when a Wi-Fi packet arrives at the states where
there is no free unlicensed band server (i.e., $D_{f} = 0$) and the unlicensed band serving packet is LAA (i.e., $x \neq 0$), this Wi-Fi packet will be dropped. Therefore,
\begin{align}
\label{block_prob_LAA}
P_{b, \ell} &= \sum_{(w, x, y, z) \in \{\mathbf{n} | z = Q, \mathbf{n} \in \mathbf{S}\}}\pi_{w, x, y, z},\\
\label{block_prob_WF}
P_{b, w} &= \sum_{(w, x, y, z) \in \{\mathbf{n} | x = D, \mathbf{n} \in \mathbf{S}\}}\pi_{w, x, y, z}.
\end{align}

The steady state probabilities $\pi_{w, x, y,
z}$ and $P_{b, \ell}$ and $P_{b, w}$ can be computed by the
iterative algorithm in~\cite{Phone2001}.

\section{Performance Evaluation}
\label{sect:performance}

In this study, the analytic models are validated by simulation experiments. We adopt the FTP traffic model~\cite{3GPP36.814}) as the packet traffic model destined to the UE.
Furthermore, the LBT-based allocation with the buffering mechanism is evaluated by
simulation experiments without analytic modeling. The model follows
the discrete event-driven simulation approach in~\cite{Phone2001}. We run $1,000,000$ application sessions in an experiment to ensure the convergence of the simulation results.
Table~\ref{analysis_sim} lists $P_{b, \ell}$ and $P_{b, w}$ values
for both analytic and simulation models for LBT-based allocation with the buffering mechanism. The details of the
parameter setup in this table will be described in the following
section. In Table~\ref{analysis_sim}, the errors between simulation and analytic
models are below $8\%$in most cases and are always less than
$6\%$. The fraction of time by the LBT mechanism induces the same deviations between analytical and simulation under low load conditions.
Since the Markov analytical model has a memoryless property, and the unsteady state is caused by the LBT mechanism, the analytical results cannot be accurately converged.
Especially, the ON/OFF state switching is more frequent, and the error is more obvious.
The inherent lack of ON/OFF unsteady state by the LBT mechanism, as well as imperfect convergence of the memoryless property, may induce some deviations between the analytical and simulation.
However, the overall trends of simulation and analysis results are still consistent.
We may further reduce the error if we consider an extreme case where we do not put the LBT into consideration, i.e., we always turn on the LAA small cell.
The validation results are listed in Table~\ref{analysis_sim_2}, which
shows that the errors between analysis and simulation results are within $1\%$.

\begin{table}[t]
\centering
\caption{Validation of the simulation and analysis results for the LAA and Wi-Fi packets.}
\resizebox{1\columnwidth}{!}{
\begin{tabular}{|c|c|c|c|c|c|c|}
\hline
\multicolumn{7}{|c|}{$E[t_{w}] = \tfrac{1}{40}~\mbox{sec}, E[t_{\ell, u}] = \tfrac{1}{25}~\mbox{sec}, E[t_{on}] = E[t_{off}] = 10~\mbox{sec}$}\\
\multicolumn{7}{|c|}{$E[t_{s}] = 1~\mbox{sec}, D=1, Q=Q_{\theta}=2$}\\
\hline
\multicolumn{2}{|c|}{$\lambda_{\ell}$ (user/sec)} &25 &37 &50 &62.5 &120\\
\hline
\multicolumn{2}{|c|}{$\lambda_{w}$ (user/sec)} &5 &5 &5 &5 &5\\
\hline
$P_{b,\ell}$ & Analytic   &0.452044 &0.558498 &0.646604 &0.707167 &0.840875\\
$P_{b,\ell}$ & Simulation &0.415108 &0.527295 &0.624808 &0.692863 &0.84164\\
$P_{b,\ell}$ & Error (\%) &8.170886 &5.586949 &3.370842 &2.022719 &0.090977\\
\hline
$P_{b,w}$ & Analytic   &0.552184 &0.657476 &0.710252 &0.734986 &0.765312\\
$P_{b,w}$ & Simulation &0.58435  &0.698707 &0.750545 &0.766701 &0.760551\\
$P_{b,w}$ & Error (\%) &5.825232 &6.271103 &5.673057 &4.315048 &0.622099\\
\hline
\end{tabular}
}
\label{analysis_sim}
\end{table}

\begin{table}[t]
\centering
\caption{Validation of the simulation and analysis results for the LTE (None of LBT) and Wi-Fi packets.}
\resizebox{1\columnwidth}{!}{
\begin{tabular}{|c|c|c|c|c|c|c|}
\hline
\multicolumn{7}{|c|}{$E[t_{w}] = \tfrac{1}{40}~\mbox{sec}, E[t_{\ell, u}] = \tfrac{1}{25}~\mbox{sec}, D=1, Q=2$}\\
\hline
\multicolumn{2}{|c|}{$\lambda_{\ell}$ (user/sec)} &25 &37 &50 &62.5 &120\\
\hline
\multicolumn{2}{|c|}{$\lambda_{w}$ (user/sec)} &5 &5 &5 &5 &5\\
\hline
$P_{b,\ell}$ & Analytic   &0.250425 &0.409601 &0.532753 &0.614984 &0.792439\\
$P_{b,\ell}$ & Simulation &0.255031 &0.412148 &0.535449 &0.616789 &0.793422\\
$P_{b,\ell}$ & Error (\%) &1.839273 &0.621825 &0.506051 &0.293504 &0.124047\\
\hline
$P_{b,w}$ & Analytic   &0.745041 &0.870437 &0.931242 &0.959145 &0.992457\\
$P_{b,w}$ & Simulation &0.743667 &0.870636 &0.929864 &0.958482 &0.99174\\
$P_{b,w}$ & Error (\%) &0.001844 &0.022862 &0.147974 &0.069124 &0.072245\\
\hline
\end{tabular}
}
\label{analysis_sim_2}
\end{table}

\begin{figure}[t]
\centering\includegraphics[width=.5\textwidth]{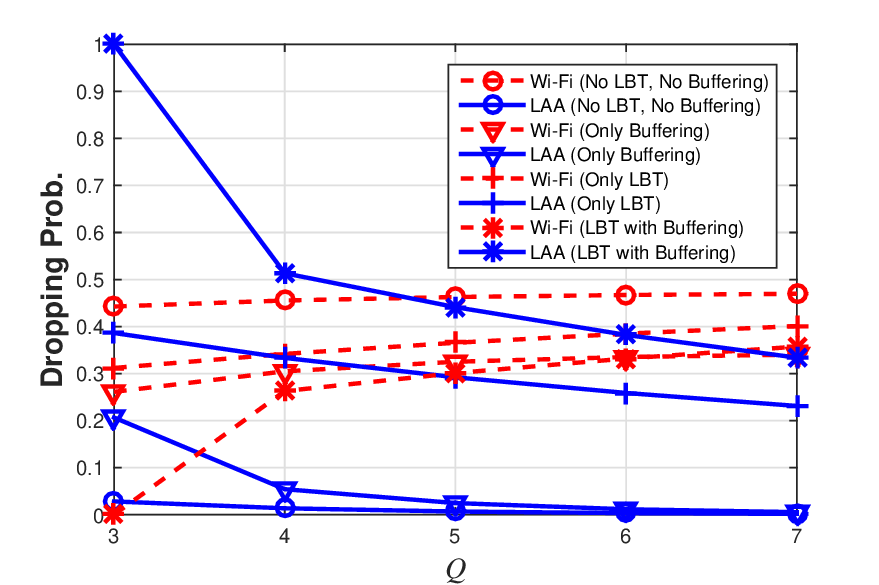}
\caption{Impact of queue size on the dropping probability.}
\label{queue_size}
\end{figure}

Fig.~\ref{queue_size} plots $P_{b, \ell}$ and $P_{b, w}$ as a function of $Q$ for the comparison of the different cases.
In the Fig.~\ref{queue_size}, we set $\lambda_{\ell} = \lambda_{w} = 0.5\mu_{w}, \mu_{\ell, u} = \mu_{w}, \mu_{on} = \mu_{off} = 0.1\mu_{w}, \mu_{s} = \mu_{w}$ and $Q_{\theta} = 2$.
Fig.~\ref{queue_size} shows that the case without LBT and buffering outperforms the other cases in terms of the $P_{b, \ell}$ performance, and the case with LBT and buffering outperforms the other cases in terms of the $P_{b, w}$ performance.
We observe that the case with buffering only outperforms the case without buffering in terms of the packet dropping probability $P_{b, w}$, which implies that the buffering mechanism for LAA packets significantly reduces the Wi-Fi packet dropping probability. When $Q$ is small, the improvements of $P_{b, \ell}$ and $P_{b, w}$ for the case with buffering only over the case with LBT only are significant. When $Q$ becomes large, to maintain both QoS for the LAA and Wi-Fi packet data users, the LBT with buffering is the better choice.

\section{Conclusion and Future Work}
\label{sect:conclusion}
This letter proposed a model including the interaction between LAA and Wi-Fi by using a global system state and a waiting queue that buffers the LAA packets when no channel is available.
We considered the LBT-based unlicensed band allocations with a buffering mechanism in the proposed analytical model.
Simulations have been used to corroborate the analytical results, as close agreement between the two is observed.
The buffering mechanism significantly reduces both the LAA and Wi-Fi packet dropping probability.
In future work, we can further extend the proposed analytical model by considering more scenarios:
1) user mobility model;
2) licensed band sharing (unlicensed band sharing) for LTE (Wi-Fi) and D2D-U communications.

\ifCLASSOPTIONcaptionsoff
  \newpage
\fi

\renewcommand\refname{Reference}
\bibliographystyle{IEEEtran}

\begin{thebibliography}{9}

\bibitem{3GPP2015}
3GPP, ``Study on licensed-assisted access to unlicensed spectrum (Release 13),'' 3GPP TR 36.889, Tech. Rep., June 2015.

\bibitem{Maule2018}
M. Maule, D. Moltchanov, P. Kustarev, M. Komarov, S. Andreev, and Y. Koucheryavy, ``Delivering fairness and QoS guarantees for LTE/Wi-Fi coexistence under LAA operation,'' \textit{IEEE Access}, vol. 6, pp. 7359--7373, Jan. 2018.

\bibitem{Tang2018}
Z. Tang, X. Zhou, Q. Hu, and G. Yu, ``Throughput analysis of LAA and Wi-Fi coexistence network with asynchronous channel access,'' \textit{IEEE Access}, vol. 6, pp. 9218--9226, Aug. 2018.

\bibitem{Song2016}
Y. Song, K. W. Sung, and Y. Han, ``Coexistence of Wi-Fi and cellular with listen-before-talk in unlicensed spectrum,'' \textit{IEEE Commun. Lett.}, vol. 20, no. 1, pp. 161--164, Jan. 2016.

\bibitem{Chen2015}
C. Chen, R. Ratasuk, and A. Ghosh, ``Downlink performance analysis of LTE and WiFi coexistence in unlicensed bands with a simple listen-before-talk scheme,'' in \textit{Proc. IEEE 81st Veh. Technol. Conf. (VTC-Spring)}, May 2015, pp. 1--5.

\bibitem{3GPP36.814}
3GPP, ``Technical specification group radio access network; further advancements for E-UTRA physical layer aspects (Release 9),'' 3GPP TR 36.814, Tech. Rep., Mar. 2010.

\bibitem{Zachary1991}
S. Zachary, ``On blocking in loss networks,'' \textit{Adv. Appl. Prob.}, vol. 23, pp. 355--372, June 1991.

\bibitem{Kelly1991}
F. P. Kelly, ``Loss networks,'' \textit{Ann. Appl. Prob.}, vol. 1, no. 3, pp. 319--378, Sept. 1991.

\bibitem{Phone2001}
P. Lin and Y.-B. Lin, ``Channel allocation for GPRS,'' \textit{IEEE Trans. Veh. Technol.}, vol. 50, no. 2, pp. 375--387, Mar. 2001.

\end{thebibliography}

\vspace{-0.4in}

\begin{IEEEbiography}[{\includegraphics[width=1in,height=1.25in,clip,keepaspectratio]{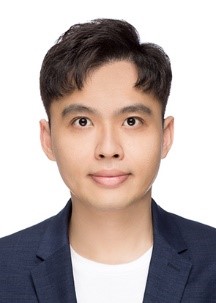}}]{Po-Heng Chou} (Member, IEEE) received the Ph.D. degree from the Graduate Institute of Communication Engineering (GICE), National Taiwan University (NTU), Taipei, Taiwan, in 2020. His research interests include AI for communications, deep learning-based signal processing, wireless networks, and wireless communications, etc.
He was a Postdoctoral Fellow at the Research Center for Information Technology Innovation (CITI), Academia Sinica, Taipei, Taiwan, from Sept. 2020 to Sept. 2024. 
He has been elected as the Distinguished Postdoctoral Scholar of CITI by Academia Sinica from Jan. 2022 to Dec. 2023. He is invited to visit Virginia Tech (VT) Research Center, Arlington, VA, USA, as a Visiting Fellow, from Aug. 2023 to Feb. 2024.
He received the Partnership Program for the Connection to the Top Labs in the World (Dragon Gate Program) from the National Science and Technology Council (NSTC) of Taiwan to perform advanced research at VT Institute for Advanced Computing, Alexandria, VA, USA, from Jan. 2025 to present.

\end{IEEEbiography}


\end{document}